\documentclass{article}

\usepackage[final]{nips_2016}

\usepackage[utf8]{inputenc} % allow utf-8 input
\usepackage[T1]{fontenc}    % use 8-bit T1 fonts
\usepackage{hyperref}       % hyperlinks
\usepackage{url}            % simple URL typesetting
\usepackage{booktabs}       % professional-quality tables
\usepackage{amsfonts}       % blackboard math symbols
\usepackage{nicefrac}       % compact symbols for 1/2, etc.
\usepackage{microtype}      % microtypography
\usepackage{graphicx}
\usepackage{algorithm,algorithmic}
\usepackage[utf8]{inputenc}
\graphicspath{{figures/}}

\usepackage{amsmath,amssymb,amsthm,mathrsfs,amsfonts,dsfont}
\newtheorem{definition}{Definition}

\graphicspath{{figures/}}

\newcommand*\samethanks[1][\value{footnote}]{\footnotemark[#1]}

\title{Convolutional Kitchen Sinks for Transcription Factor Binding Site Prediction}

\author{
  Alyssa Morrow\thanks{These authors contributed equally} \\
  \And
  Vaishaal Shankar\samethanks \\
  \AND
  Devin Petersohn, Anthony Joseph, Benjamin Recht, Nir Yosef \\
  University of California, Berkeley \\
  \{akmorrow,shankar,devin.petersohn,adj,brecht,niryosef\}@berkeley.edu \\
}

\begin{document}

\maketitle
\begin{abstract}
We present a simple and efficient method for prediction of transcription factor binding sites from DNA sequence. Our method computes a random approximation of a convolutional kernel feature map from DNA sequence and then learns a linear
model from the approximated feature map. Our method outperforms state-of-the-art deep learning methods on five out of six test datasets from the ENCODE consortium, while training in less than one eighth the time.
\end{abstract}

\section{Introduction}
Understanding binding affinity between proteins and DNA sequence is a crucial step in deciphering
the regulation of gene expression. Specifically, characterizing the binding affinity of transcription factor proteins (TFs) to DNA sequence determines the relative expression of genes downstream from a TF binding site.

The recent advent of sequencing technologies,
such as chromatin immunoprecipitation with massively parallel DNA sequencing (ChIP-seq),
provides us with genome-wide binding specificities for 187 TFs across
98 cellular contexts of interest from the ENCODE consortium \cite{encode2004encode}. These specificities
can be thresholded to define high-confidence bound and unbound regions for a given TF. Given the location of these binding sites, we can formulate a binary sequence classification problem, classifying regions bound and unbound by a TF as positive and negative, respectively.
Using a binary sequence classification model, we can predict binding sites in new cellular contexts,
learning regulatory behavior without the expense of ChIP-seq experiments.

String kernel methods are well understood and have been extensively used for sequence classification
\cite{jaakkola1999using, eskin2002mismatch, leslie2002spectrum}.
Specifically, Fletez-Brant et al. and Lee et al. \cite{fletez2013kmer, lee2015method} have applied string kernel methods
to the prediction of transcription factor binding sites.
However, kernel methods require pairwise comparison
between all $n$ training sequences and thus incur an expensive $\mathcal{O}(n^{2})$ computational and storage complexity,
making them computationally intractable for large data sets.

Recently, convolutional neural networks (CNN) have been successful for prediction of
TF binding sites \cite{zhou2015predicting,  alipanahi2015predicting, kelley2016basset}.
CNNs generalize well by encoding spatial invariance during training.
Fast convolutions on a Graphical Processing Unit (GPU) allows CNNs to train on large datasets.
However, the actual design of the neural network greatly impacts model performance,
yet there is no clear understanding of how to design a network for a particular task.
Furthermore there is no generally accepted network architecture for the task of TF binding
site prediction from DNA sequence.

In this work, we present a convolutional kernel approximation algorithm that maintains the spatial invariance
and computational efficiency of CNNs.
Dubbed Convolutional Kitchen Sinks (CKS), our algorithm learns a model from the output of a
1 layer random convolutional neural network \cite{rahimi2009weighted}. All the parameters of the network are independent and identically distributed (IID)
random samples from a gaussian distribution with a specified variance.
We then train a linear model on the output of this network.
Our results show that for five out of six transcription factors, CKS outperform
current state-of-the art CNN implementations, while maintaining a simple architecture and training eight times faster than a CNN.

\section{Method}
The task of transcription factor (TF) binding site prediction from DNA sequence reduces to binary sequence classification.
We present a randomized algorithm for finding an embedding of sequence data apt for linear classification (Algorithm \ref{alg:cks}). Our algorithm is closely related to the work of convolutional kernel networks, which approximates a convolutional kernel feature map via a nonconvex optimization objective \cite{mairal2014convolutional}. However, unlike Mairal et al. \cite{mairal2014convolutional},
we approximate the convolutional kernel feature map via random projections in the style of  Rahimi et al. \cite{rahimi2009weighted, rahimi2007random}.

We will first define the convolutional $n$-gram kernel, and then analyze why it has
desired properties for the task of string classification.
Note that we use the term $n$-gram to refer to a contiguous sequence of $n$ characters,
whereas computational biology literature refers to the same concept as a $k$-mer.

\begin{definition}[Convolutional $n$-gram kernel]
Let $x,y$ be strings of length $d$ from an underlying alphabet $\mathcal{A}$, and let $\mathds{H}(x,y)$ denote the Hamming distance between the two strings. Let $x_{i:j}$ denote the substring of $x$ from index $i$ to $j - 1$. Let $n$ be an integer less than $d$ and let $\gamma$ be a real valued positive number denoting the width of the kernel. The kernel function $K_{n,\gamma}(x, y)$ is defined as:

\vspace{-0.4cm}
\begin{equation}
K_{n,\gamma}(x, y) = \displaystyle\sum_{i=0}^{d - n} \displaystyle\sum_{j=0}^{d - n}
\exp(-\gamma \mathds{H}^{2}(x_{i:i+n}, y_{j:j+n}))
\label{eq:convkernel}
\end{equation}
\end{definition}

To gain intuition for the behavior of this kernel, take $\gamma$ to be a large value. It follows that
$\exp(-\gamma \mathds{H}^{2}(x_{i:i+n}, y_{j:j+n})) \approx \mathds{1}[x_{i:i+n} = y_{j:j+n}]$.

This combinatorial reformulation results in the following well studied Spectrum Kernel (Definition~\ref{def:spectrum}).

\begin{definition}[Spectrum Kernel]
\label{def:spectrum}
Let $\mathcal{S}_{n}(\mathcal{A})$ be the set of all length $n$ contiguous substrings in $\mathcal{A}$, and $\#(x,s)$ count the occurrences of $s \in x$ \cite{leslie2002spectrum}.

\vspace{-0.4cm}
\begin{equation}
K_{spec}(x, y) = \displaystyle\sum_{s \in \mathcal{S}_n(\mathcal{A})} \#(x,s)\#(y,s)
\label{eq:spectrum}
\end{equation}
\end{definition}

\begin{algorithm}[b]
   \caption{Convolutional Kitchen Sink for sequences}\label{alg:cks}
   \begin{algorithmic}[1]
      \INPUT $x_{i} \ldots x_{N} \in \mathds{R}^{d}$ (input sequences), $\gamma$ (width of kernel),  $n$ (convolution size), $M$ (the approximation dimension, number of kitchen sinks)
      \FOR{$j \in \left\{0 \ldots M \right\}$}
         \STATE
         $w_{j} \sim \mathcal{N}(0, \gamma I_{n})$ \hspace*{2cm} \textit{Sample kitchen sink from gaussian}

         \STATE
         $b_{j} \sim U(0, 2\pi)$ \hspace*{2.28cm} \textit{Sample phase from uniform disk} \\

        \FOR{$i \in \left\{0 \ldots N \right\}$}
             \STATE
             $z_{ij} = w_{j} \ast x_{i}$ \hspace*{2.0cm} \textit{Convolve filter with input sequence} \label{cks:line5} \\
             \STATE
             $c_{ij} = \cos(z_{ij} + b_{j})$ \hspace*{1.3cm}\textit{Add phase and compute element-wise cosine} \\
            \hspace*{4.0cm} \textit{Note} $z_{ij}$ and $c_{ij}$ \textit{are vectors in } $\mathcal{R}^{d - n + 1}$ \label{cks:line5} \\

             \STATE $\phi(x_{i})_{j} = \sqrt{\frac{2}{M}} \displaystyle \sum_{k=0}^{d - n} c_{ijk} $ \hspace*{0.4cm} \textit{Average to get the $j$th output feature value for sequence $x_{i}$} \\
            \ENDFOR
     \ENDFOR
      \OUTPUT $\phi(x_{i}) \ldots \phi(x_{N})$
   \end{algorithmic}
\end{algorithm}

Other string kernel methods such as the mismatch \cite{eskin2002mismatch} and gapped $n$-gram kernel \cite{ghandi2014enhanced} allow for partial mismatches between $n$-grams. We note that decreasing $\gamma$ in Equation~\ref{eq:convkernel} relaxes the penalty of $n$-gram mismatches between disappoints, thereby capturing the behavior  of the mismatch and gapped $n$-gram kernels \cite{eskin2002mismatch, ghandi2014enhanced}. Note that Equation~\ref{eq:convkernel} is computationally prohibitive, as it takes $\Omega(nd^{2})$ to compute each of the $N^{2}$ entries in the kernel matrix.
Furthermore, the feature map induced by the kernel in Equation~\ref{eq:convkernel} is infinite dimensional, so the kernel matrix is necessary.

Instead, we turn to a random approximation of Equation~\ref{eq:convkernel} (see Algorithm \ref{alg:cks}).
Since our kernel is a sum of non linear functions it suffices to define a feature map $\hat{\phi}$ on sequences x and y that approximates each term in the sum from Equation~\ref{eq:convkernel}:

\vspace{-0.4cm}
\begin{equation}
\label{eq:approx}
\exp(-\gamma \mathds{H}(x_{i:i+n}, y_{j:j+n})) \approx \hat{\phi}(x_{i:i+n})^{T}\hat{\phi}(y_{j:j+n})
\end{equation}

Claim 1 from Rahimi et al. \cite{rahimi2007random} states that for $j \in \left\{0 \ldots M - 1\right\}$, if we choose $\hat{\phi}(x_{i:i+n})_{j} = \sqrt{\frac{2}{M}} \cos(w_{j}^{T}x_{i:i+n} + b_{j})$, where $w_{j} \sim \mathcal{N}(0, \gamma)$, $b_{j} \sim U(0, 2\pi)$, then $\hat{\phi}(x_{i:i+n})$ satisfies Equation~\ref{eq:approx}.
Note that to use Claim 1, we represent Hamming distance in Equation~\ref{eq:convkernel} as an L2 distance. We refer to each $w_{j}$ as a ``random kitchen sink".
The result in in Rahimi et al. \cite{rahimi2007random} (Claim 1) gives strong guarantees that $\hat{\phi}(x)^{T}\hat{\phi}(y)$ concentrates exponentially fast to Equation~\ref{eq:convkernel}, which means we can set $M$, the number of kitchen sinks, to be small.

Algorithm~\ref{alg:cks} details the kernel approximation. Note that in Algorithm~\ref{alg:cks}, line~\ref{cks:line5} we reuse
$w_{j}$ across all $x_{i:i+n}$ in Equation~\ref{eq:convkernel} by a convolution.
Algorithm~\ref{alg:cks} is a finite dimensional approximation of the feature map induced by the kernel in Equation~\ref{eq:convkernel} directly,
circumventing the need for a kernel matrix.
The computational complexity of Algorithm~\ref{alg:cks} is $\mathcal{O}(NMdn)$.

For the task of TF binding site prediction we let alphabet $\mathcal{A} = \left\{A,T,C,G\right\}$, and set $n=8$, similar to common parameter configuration for DNA sequence \cite{fletez2013kmer, alipanahi2015predicting, ghandi2014enhanced}.

\section{Results}
% table for deepbinds data

\begin{table}[b]
\centering
% \vspace{-0.4cm}
\caption{Comparison of ROC Area under Curve values (AUC) between DeepBind and CKS tested on 500 bound regions from ENCODE and 500 synthetic unbound regions.}
\label{table:eval}
\hspace*{-0.96cm}
\begin{tabular}{|l|l|l|l|l|l|l|}
\hline

\textbf{TF} & \textbf{\begin{tabular}[c]{@{}l@{}}Train\\ Cell Type\end{tabular}} & \textbf{\begin{tabular}[c]{@{}l@{}}Test\\ Cell Type\end{tabular}} & \textbf{\begin{tabular}[c]{@{}l@{}}Train\\ Size\end{tabular}} & \textbf{\begin{tabular}[c]{@{}l@{}}Train\\ Time\end{tabular}} & \textbf{\begin{tabular}[c]{@{}l@{}}DeepBind\\ AUC\end{tabular}} & \textbf{\begin{tabular}[c]{@{}l@{}}CKS\\ AUC\end{tabular}} \\ \hline
ATF2                & H1-hESC                    & GM12878                          & 10998   & 154s      & 0.72                     & 0.77  \\ \hline
ATF3                & H1-hESC                    & HepG2                            & 8616    & 139s      & 0.94                     & 0.95  \\ \hline
ATF3                & H1-hESC                    & K562                              & 8616    & 139s     & 0.83                     & 0.84   \\ \hline
CEBPB               & HeLa-S3                    & A549                            & 121010  & 1620s      & 0.99                     & 0.99      \\ \hline
CEBPB               & HeLa-S3                    & K562                            & 121010  & 1620s      & 0.99                     & 0.98       \\ \hline
EGR1               & K562                        & GM12878                        & 72996   & 772s        & 0.94                     & 0.96 \\ \hline
EGR1               & K562                        & H1-hESC                        & 72996   & 772s       & 0.87                     & 0.92  \\ \hline
EP300              & HepG2                       & SK-N-SH                         & 54828   & 519s       & 0.67                     & 0.70  \\ \hline
EP300              & HepG2                       & K562                            & 54828   & 519s       & 0.66                     & 0.81 \\ \hline
STAT5A             & GM12878                     & K562                             & 13846   & 199s      & 0.65                     & 0.79  \\ \hline

\end{tabular}
\label{table:deepbind}
\end{table}

\begin{table}[b]
\centering
\vspace{-0.1cm}
\caption{Comparison of ROC Area under Curve values (AUC) between DeepBind and CKS tested on 100,000 bound and unbound regions from ENCODE. Because both experiments trained on the same dataset, the train cell types, train times, and train sizes are the same as in Table ~\ref{table:deepbind}.}
\label{table:test}
\hspace*{-1cm}
\begin{tabular}{|l|l|l|l|}
\hline
\textbf{TF} & \textbf{\begin{tabular}[c]{@{}l@{}}Test\\ Cell Type\end{tabular}} & \textbf{\begin{tabular}[c]{@{}l@{}}DeepBind\\ AUC\end{tabular}} & \textbf{\begin{tabular}[c]{@{}l@{}}CKS\\ AUC\end{tabular}} \\ \hline
ATF2                                       & GM12878                        & 0.56                     & 0.57           \\ \hline
ATF2                                       & MCF7                           & 0.93                     & 0.76           \\ \hline
EGR1                                       & GM12878                        & 0.87                     & 0.91           \\ \hline
EGR1                                       & H1-hESC                        & 0.77                     & 0.85           \\ \hline
EGR1                                       & HCT116                         & 0.77                     & 0.82           \\ \hline
EGR1                                       & MCF7                           & 0.84                     & 0.86           \\ \hline
\end{tabular}
\label{table:encode}
\end{table}

\begin{figure}
\label{fig:encode}
\begin{tabular}{cccc}
  \hspace{-2.2cm}
  \includegraphics[width=0.3\textwidth]{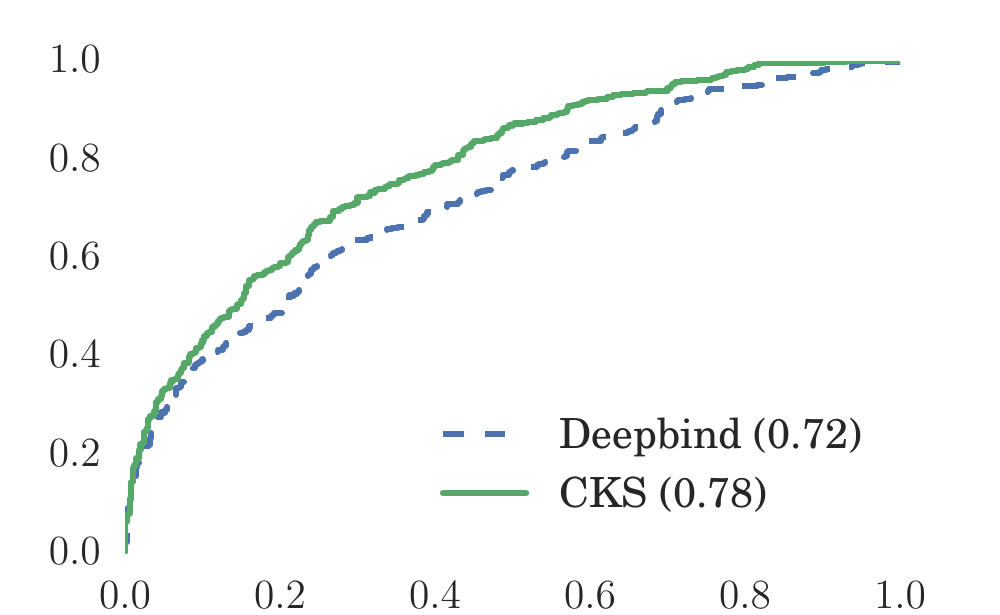}  &
  \includegraphics[width=0.3\textwidth]{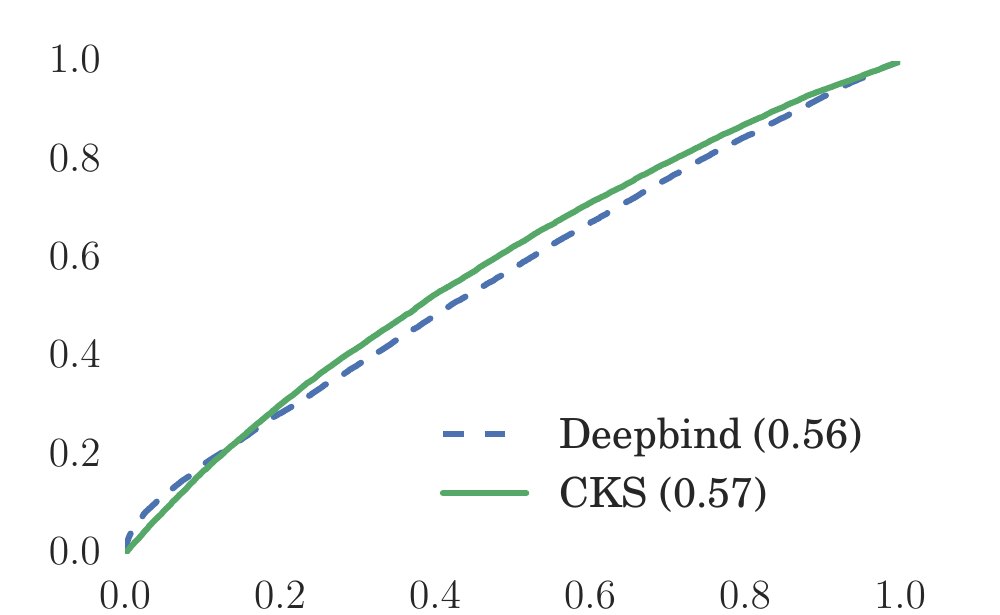} &
  \includegraphics[width=0.3\textwidth]{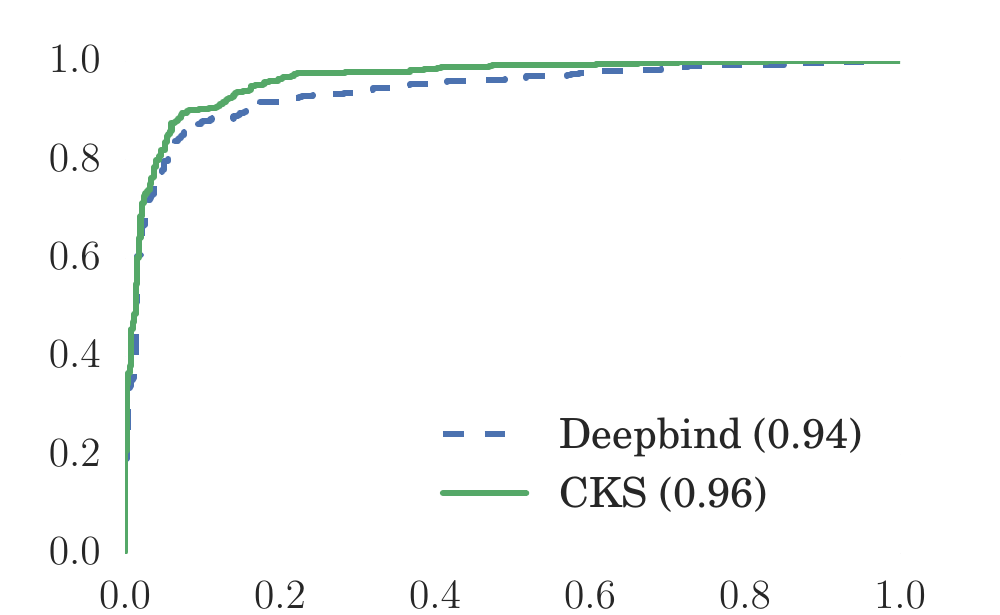} &
  \includegraphics[width=0.3\textwidth]{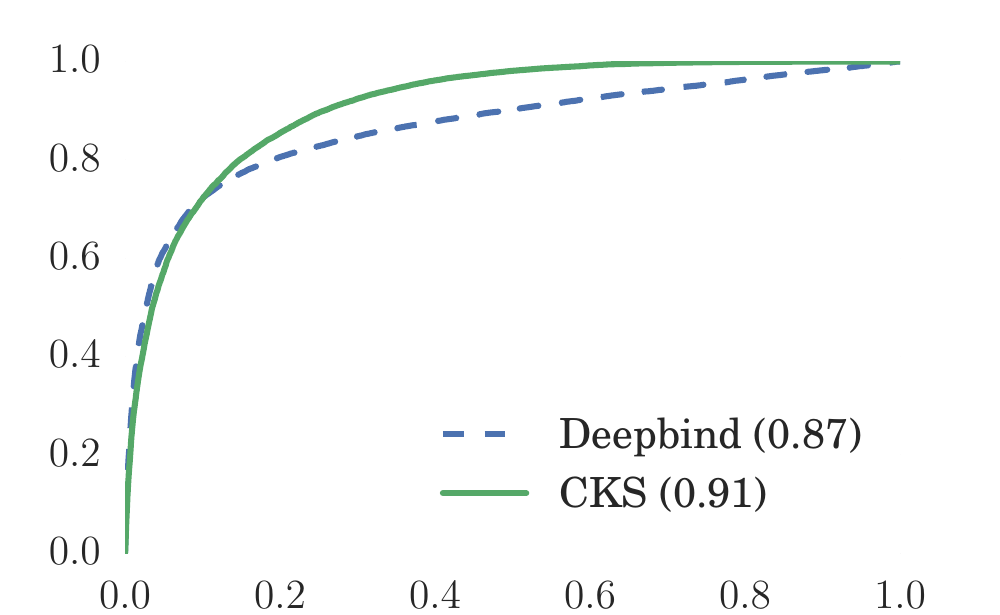} \\
  A) & B) & C) & D)
  \end{tabular}
  \caption{ROC of DeepBind and CKS for TFs EGR1 and ATF2 for GM12878. A) ROC for ATF2 on DeepBind's test set. B) ROC for ATF2 on the ENCODE set. C) ROC for EGR1 on DeepBind's test set. D) ROC for EGR1 on the ENCODE set.}
\end{figure}

We compare our CKS to DeepBind, a state-of-the-art CNN approach for predicting transcription factor (TF) binding sites. We compare to DeepBind over other CNN methods \cite{zhou2015predicting, kelley2016basset} due to its primary attention to DNA sequence specificity and ability to identify fine grained (101 bp) locations of binding affinity.

\subsection{Datasets}
We train and evalute on datasets preprocessed from the ENCODE consortium.
Because binding affinity is TF specific, we use separate train and evaluation sets for each TF.

We use the same training sets as DeepBind's publically available models. We then evaluate on DeepBind's
test sets as well as a larger dataset processed directly from ENCODE.

DeepBind's test sets consist of 1000 regions for each cell type over six TFs. Each set consists of
500 positive sequences extracted from regions of high ChIP-seq signal and 500 synthetic negative sequences generated from dinucleotide shuffle of positive sequences \cite{alipanahi2015predicting}.

The second test dataset consists of 100,000 regions extracted from ChIP-seq datasets for TFs ATF2 and EGR1 across multiple cell types. Positive sequences are extracted from regions of high ChIP-seq signal.
Negative sequences are extracted from regions of low ChIP-seq signal with exposed chromatin.

\subsection{Experimental Setup}
Experiments for DeepBind and CKS were run on one machine with 24 Xeon processors, and 256 GB of ram and 1 Nvidia Tesla K20c GPU.

We train a linear model minimizing squared loss with an L2 penalty of $\lambda$ on the output of the CKS defined in Algorithm~\ref{alg:cks}.
We do not tune the hyper-parameters $n$ (convolution size)
and $M$ (number of kitchen sinks), and leave them constant at $8$ and $8192$ respectively.
We tune the hyper paraemters $\gamma$ (kernel width) and $\lambda$ on held out data from the train set.
To assess generalization across cellular contexts, we train and evaluate on separate cell types.

\subsection{Evaluation}
We compare DeepBind against CKS using area under the curve (AUC) of Receiver Operating Characteristic (ROC).
We choose AUC as a metric for binary classifcation due to its ability to measure both TF binding site detection and false positive rates.

We detail our experimental results and compare to DeepBind's pretrained models in Tables \ref{table:deepbind} and \ref{table:encode}.
We also show ROCs for ATF2 and EGR1 on both datasets in Figure~\ref{fig:encode}.

Our AUC is competitive (within 0.01) or superior to that of DeepBind except for ATF2 on MCF7 cell type.
Furthermore on five out of six large ENCODE test sets, our AUC is strictly greater than DeepBind.

We measure DeepBind's training time on TF EGR1, trained on K562 with $72,996$ train sequences.
DeepBind's training procedure takes $6497$ seconds to learn $2123$ parameters.
For comparison, training time for CKS takes $712$ seconds (Table \ref{table:deepbind}) to learn $16384$ parameters, which is approximately eight times faster than DeepBind's runtime.

\section{Conclusion and Future Work}
In this paper, we show that Convolutional Kitchen Sinks train eight times faster and has superior predictive
performance to CNNs.
We note that our current work focuses on binding affinity in the context of DNA sequence,
making this model agnostic to specific cell contexts of interest.
Because Algorithm 1 is not specific to DNA sequence,
positional counts of chromatin
accessibility and gene expression data can be aggregated with current implementation
to account for cell type specific information. We leave this extension for future work.

\bibliographystyle{unsrt}
\bibliography{main}

\end{document}